\documentclass[aps,prl,twocolumn,letterpaper,superscriptaddress]{revtex4}
\usepackage{amssymb}
\usepackage{amsmath}
\usepackage{graphicx}
\usepackage{hyperref}
\def\rd{R_{\mathrm{D}}}

\def\vd{V_{\mathrm{D}}}
\def\Idc{I_{\mathrm{DC}}}
\def\Imax{I_{\mathrm{max}}}
\def\Ihold{I_{\mathrm{hold}}}
\def\nfield{B_{\mathrm{N}}}

\def\um{\mathrm{\mu m}}

\def\perm{\mathrm{m}^{-2}}
\def\mob{\mathrm{m}^{2}}

\def\vl{V_{\mathrm{L2}}}
\def\fnmr{f_{\mathrm{NMR}}}
\def\vqpc{\nu_{\mathrm{D}}}
\def\vcf{\nu_{\mathrm{CF}}}

\begin{document}
\title{Dynamic Nuclear Polarization in the Fractional Quantum Hall Regime}
\author{A.~Kou}
\affiliation{Department of Physics, Harvard University, Cambridge, Massachusetts
02138, USA}

\author{D.~T.~McClure}
\affiliation{Department of Physics, Harvard University, Cambridge, Massachusetts
02138, USA}

\author{C.~M.~Marcus}
\affiliation{Department of Physics, Harvard University, Cambridge, Massachusetts
02138, USA}

\author{L.~N.~Pfeiffer}
\affiliation{Department of Electrical Engineering, Princeton University, Princeton, New
Jersey 08544, USA}

\author{K.~W.~West}
\affiliation{Department of Electrical Engineering, Princeton University, Princeton, New
Jersey 08544, USA}
\date{\today}

\begin{abstract}
We investigate dynamic nuclear polarization in quantum point contacts (QPCs) in the integer and fractional quantum Hall regimes. Following the application of a dc bias, fractional plateaus in the QPC shift symmetrically about half filling of the lowest Landau level, $\nu=1/2$, suggesting an interpretation in terms of composite fermions. Polarizing and detecting at different filling factors indicates that Zeeman energy is reduced by the induced nuclear polarization. Mapping effects from integer to fractional regimes extends the composite fermion picture to include hyperfine coupling. 
\end{abstract}

\maketitle

An appealing physical picture of the fractional quantum Hall (FQH) effect is the composite fermion  model \cite{Jain07}, in which an even number, $2m$, of flux quanta ($\phi_{0} = h/e$) bind to each electron, creating a composite fermion (CF) that feels an effective field, $B$*$=B-2mn\phi _{0}$, where $B$ is the applied field perpendicular to the plane of the electron gas, and $n$ is the electron density. The effective field quantizes the CF energy spectrum into the analogue of electronic Landau levels; the FQH effect becomes the integer quantum Hall (IQH) effect of CFs.  At filling factor $\nu=1/2$, corresponding to $B$*$=0$ for CFs with two attached flux quanta ($m=1$), the CFs form a Fermi sea that can have ground states with different degrees of spin polarization \cite{Kukushkin:1999p6093,Freytag:2002p6482}. Composite fermions at other filling factors also have non-trivial spin-polarized ground states.  For example, $\nu=2/3$ ($\nu_{CF}=-2$) and $\nu=2/5$ ($\nu_{CF}=2$) have been observed to have both spin-polarized and spin-unpolarized ground states \cite{Eisenstein:1990p5954,Kang:1997p6459}.  

Dynamic nuclear polarization (DNP) has been used to investigate both the IQH and the FQH regime using transport measurements \cite{Wald:1994p4488,Dixon:1997p4492,lispin,Kronmueller:1999p4493,Smet:2001p5728,Smet:2002p4490,Machida:2002p5588,Freytag:2002p6482,Machida:2003p4878,Stern:2004p5798,Yusa:2005p4486,Kawamura:2007p4903,Tracy:2007p5947,Kawamura:2008p5462,Li:2009p6143,Kawamura:2009p4483,Dean:2009p8309}.  In both regimes, electron spin flips are accompanied by opposite nuclear spin flops. For instance, in gate-confined GaAs microstructures in the IQH regime, Wald {\it et al.}~\cite{Wald:1994p4488} found that scattering from the lowest (spin-up) Landau level to the second (spin-down) Landau level flops a nuclear spin from down to up, which in turn increments the Overhauser field, $\nfield$, opposite to the applied field, $B$. The resulting reduction in total effective Zeeman field was then detectable in transport through a quantum point contact (QPC) \cite{Wald:1994p4488}.  Dynamic nuclear polarization in a QPC with only Zeeman splitting has also been observed \cite{Ren:2009p8310}. In bulk two-dimensional (2D) geometries, breakdown of the IQH and FQH effects at high bias can also induce DNP \cite{Machida:2002p5588,Machida:2003p4878,Kawamura:2007p4903,Kawamura:2008p5462,Kawamura:2009p4483,Dean:2009p8309}. In this case, the direction of the resulting Overhauser field depends on experimental details. In the FQH regime, much of the work---in bulk and in microstructures---has focused on $\nu=2/3$  \cite{Kronmueller:1999p4493,Smet:2001p5728,Smet:2002p4490,Yusa:2005p4486}, where DNP is attributed to spin-flip tunneling between spin-unpolarized and spin-polarized domains. Bulk 2D studies using DNP and nuclear relaxation at $\nu=1/2$ were used to investigate the degree of spin polarization of the metallic CF state as a function of applied field \cite{Freytag:2002p6482,Tracy:2007p5947,Li:2009p6143}.  Despite the extensive literature on this topic, on both bulk and confined devices, no explicit connection between DNP in the IQH and FQH regimes---creation or detection---has been drawn to our knowledge.

\begin{figure}[b!]
\center \label{figure1}
\includegraphics[width=3in]{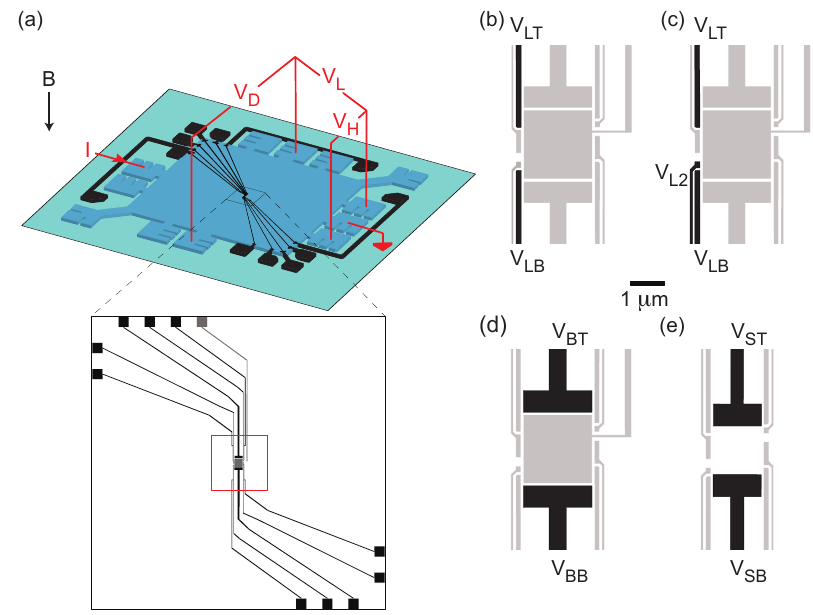}
\caption{\footnotesize{(a) Device layout, showing leads used for current bias $I$ and diagonal voltage, $\vqpc$, as well as bulk longitudinal and Hall voltages. (b-e)Gate layouts for $1~\um$, 750~nm, $2~\um$, and $1.4~\um$ constrictions, indicating depleted (black) and grounded (gray) gates.}}
\end{figure}

In this Letter, we experimentally investigate DNP in a gate-defined QPC, and identify surprising correspondences between the IQH and FQH regimes, which we then interpret within a composite fermion picture. In contrast to the situation in bulk FQH systems, where DNP may change the spin configuration at a given filling factor, we find that DNP in the vicinity of a QPC can evidently induce changes in density (hence local filling factor) within the constriction. Resistance plateaus as a function of $B$ in both IQH and FQH regimes shift and change in length following application of a nonzero dc bias. Using resistively detected nuclear magnetic resonance (NMR), we demonstrate that the applied bias induces nuclear polarization. Interestingly, the pattern of shifting plateaus is symmetric about the the half-filled first Landau level, $\nu=1/2$. Comparable shifts are also found in the IQH regime.  We then determine the sign of the induced Overhauser field to be opposed to the applied field in all cases, and estimate the magnitude of the Overhauser field by observing its effects at large filling factors, where the Overhauser field can exceed the applied field and effectively reverse the sign of the Zeeman field. Finally, we interpret related DNP effects in the IQH and FQH regimes in terms of simple Zeeman-split CF edge states. 

Measurements were carried out on four devices [Figs.~1(b-e), black gates depleted], all showing similar behavior. Data presented are from devices in Figs.~1(b,c).  The devices were fabricated on a two-dimensional electron gas (2DEG) in a symmetrically Si-doped GaAs/AlGaAs  48~nm quantum well structure located 400~nm below the wafer surface with density $n$~=~$7.8\times10^{14}~\perm$ and mobility $\mu$~=~1,300 $\mob$/V$\cdot$s measured in the dark. Similar behavior was observed on a different wafer with roughly twice the density. Square mesas were wet-etched [Fig.~1(a)], and Ti/Au (5 nm/15 nm) surface gates were patterned using electron-beam lithography. Depleted gates except $\vl$ were set to $\sim-1.5~\mathrm{V}$. Gate $\vl$, when used, was set to $\sim-0.8~\mathrm{V}$. Other gates were grounded.

Measurements were made using a current bias $I$, with dc component, $\Idc$, up to 100~nA and ac component 400~pA at 153~Hz. The electron temperature was $\sim$50~mK. We typically measure the diagonal voltage, $\vd$, which is the voltage difference between incoming edge states on opposite sides of the QPC.  Lock-in measurements of the diagonal resistance, $\rd~\equiv~d\vd/dI$, were used to determine the local filling factor in the QPC, $\vqpc \equiv h/\rd e^2$.  Two procedures were used to apply $\Idc$ to the QPC.  In the first procedure (``holding"), $\Idc$ was set to a value $\Ihold$ for a time $t_{\mathrm{hold}}$ before being set back to 0.  Unlike in less symmetric geometries \cite{Machida:2002p5588}, results did not depend on the sign of $\Ihold$.  In the second procedure (``sweeping"), $\Idc$ was swept from a positive value ($\Imax$) to a negative value ($-\Imax$) and then swept back to 0; sweep direction made no difference. The two procedures lead to similar overall behavior, as well.

Figure~2(a) shows $\rd$ as a function of $B$ and $\Idc$ in the 750 nm constriction [Fig.~1(c)], acquired using the sweeping procedure at each field then stepping the field downward.  Comparing zero-bias data taken prior to sweeping (red) with the zero-bias cut through data (black) shows that sweeping causes the $\vqpc=1/3$ plateau to extend to lower field, just past the high-field edge of the $\vqpc=2/5$ plateau in (red) data taken prior to sweeping  $\Idc$ (red).  The transition region between $\vqpc=1/3$ and $\vqpc=2/5$ becomes abrupt after sweeping dc bias; in the prior data, the transition is seen to be gradual.  Figure 2(b) shows similar extensions of plateaus for $\vqpc$ between $2/5$ and $2/3$ in the $1~\um$ constriction [Fig.~1(b)]. Here, the bias was applied using the holding procedure at each field, then the field stepped downward. The black trace shows $\rd$ immediately after the return to zero dc bias at each field while the red trace was measured with no dc bias applied. 

The pattern of shifts and extensions of plateaus of $\rd$ [Fig.~2(b)] exhibits a striking symmetry about $\vqpc=1/2$: applying then removing dc bias at each field causes all plateaus to shift toward $\vqpc=1/2$, which, as a symmetry point, does not change position \cite{Detail2}. Bulk Hall and longitudinal resistances do not exhibit any change in behavior after applying $\Idc$. 

\begin{figure}[t]
\center \label{figure2}
\includegraphics[width=3in]{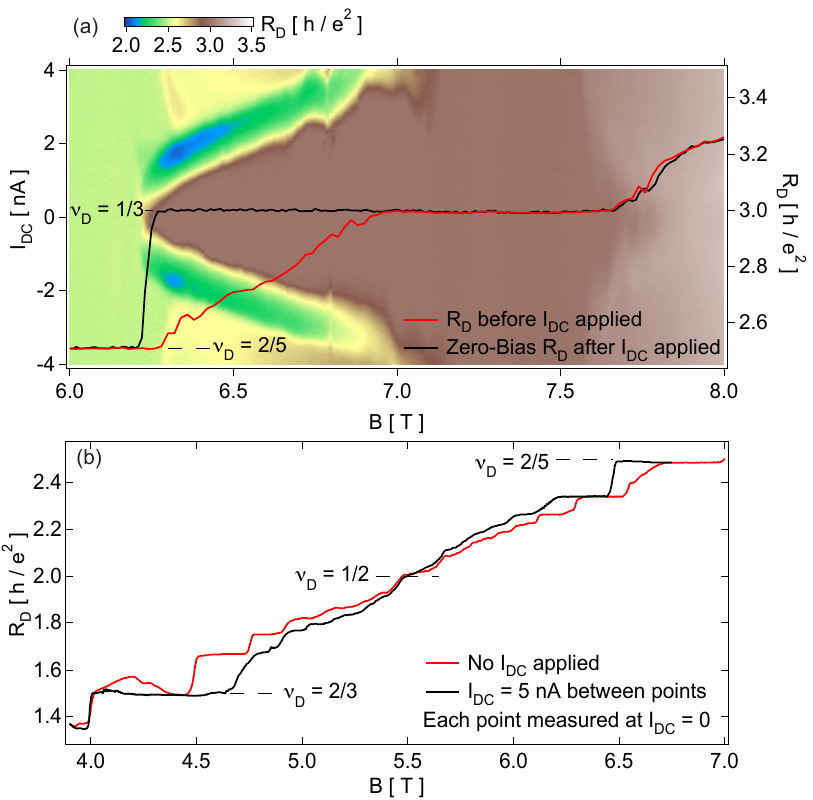}
\caption{\footnotesize{(a) Diagonal resistance of the QPC, $\rd$, (Colorscale) as a function of $B$ and $\Idc$ (left scale) in the $750$~nm constriction. Traces (right scale) show $\rd$ at $\Idc = 0$ as a function of $B$ before (red) and after (black) application of $\Idc$. Black trace is the zero-bias cut through the color plot. (b) $\rd$ as a function of $B$ in the $1~\um$ constriction before (red) and after (black) application of $\Idc$~=~5~nA.}}
\end{figure}

Similar shifts and extensions of plateaus occur when the bias is applied at a different filling factor from where its effects are observed. Changes in filling factor can be accomplished by either changing field or QPC gate voltage. This is illustrated in Figs.~3(a-c), which show $\rd$ as a function of time, measured at the same field and gate settings, following application of $\Idc$ at three different filling factors. Field and gate voltages were first set to give a well-quantized $\vqpc=3/5$ plateau in the $1~\um$ constriction prior to application of $\Idc$. Then, either field or gate voltage was used to change $\vqpc$, where dc bias sweeping procedure was applied. Field or gate voltage values were then returned to the settings where $\vqpc=3/5$ was originally observed.  In all cases---regardless of where $\Idc$ is applied---after a transient (due to residual heating from $\Idc$) $\rd$ settles at a value indicating $\vqpc=2/3$ for tens of minutes before suddenly returning to its original $\vqpc=3/5$ value. 

Plateau shifting with characteristic symmetry about $\nu=1/2$ is also observed when $\Idc$ is applied at a single filling factor rather than at each value of $B$. In Fig.~3(d), the $\Idc$ sweeping procedure applied once, just below $\nu=1/3$ ($B=7.5$~T), before sweeping field downward to $\nu=2/3$ ($B=4.0$~T) with $\Idc$ at zero. Symmetry about $\nu=1/2$ is evident despite the asymmetry of where the dc bias was applied. Similar behavior is seen when $\Idc$ is applied once at $\nu=3/5$ ($B=4.5$~T) before sweeping the magnetic field upward toward $\nu=1/3$ ($B=7.8$~T) with $\Idc = 0$ [Fig.~3(d)]. From these data, we conclude that the observed symmetry about $\nu=1/2$ reflects the {\it response} of the system to a common, roughly field-independent, physical mechanism.

\begin{figure} [t]
\center \label{figure3}
\includegraphics[width=3in]{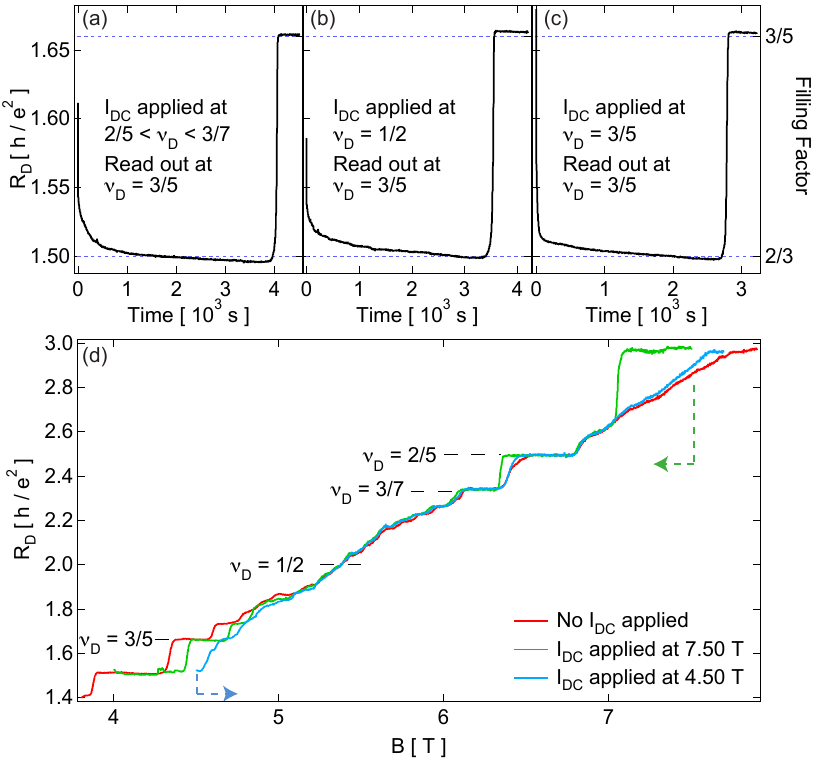}
\caption{\footnotesize{(a-c) $\rd$ as a function of time measured in the $1~\um$ constriction at $\vqpc=3/5$ after $\Idc$~=~10~nA was applied at (a) $2/5~\textless~\vqpc~\textless~3/7$, (b) $\nu=1/2$, and (c) $\vqpc=3/5$. (d) $\rd$ as a function of $B$ in the $1~\um$ constriction. Red trace indicates $\rd$ before $\Idc$ has been applied. Green trace indicates $\rd$ after $\Idc=20$~nA has been applied at $B=7.50$~T. Blue trace indicates $\rd$ after $\Idc=34$~nA has been applied at $B=4.50$~T.}}
\end{figure} 

The slow relaxation seen in Figs.~3(a-c) suggests DNP as the origin of the effects of applied bias. This is confirmed using resistively detected NMR. Following sweeping application of $\Idc$, an ac magnetic field pulse at frequency $\fnmr$ is applied using a six-turn coil that orients the ac field predominantly in the plane of the electron gas. When $\fnmr$ matches one of the expected NMR frequencies, $\rd$ returns to the value measured before applying $\Idc$. Figures~4(a,b) show depolarization signatures in $\rd$ for $^{75}$As NMR at $\vqpc=1/3$ and $\vqpc=3/5$.  Similar signatures are also observed for $^{69}$Ga and $^{71}$Ga NMR frequencies (not shown).  

\begin{figure} [t]
\center \label{figure4}
\includegraphics[width=3in]{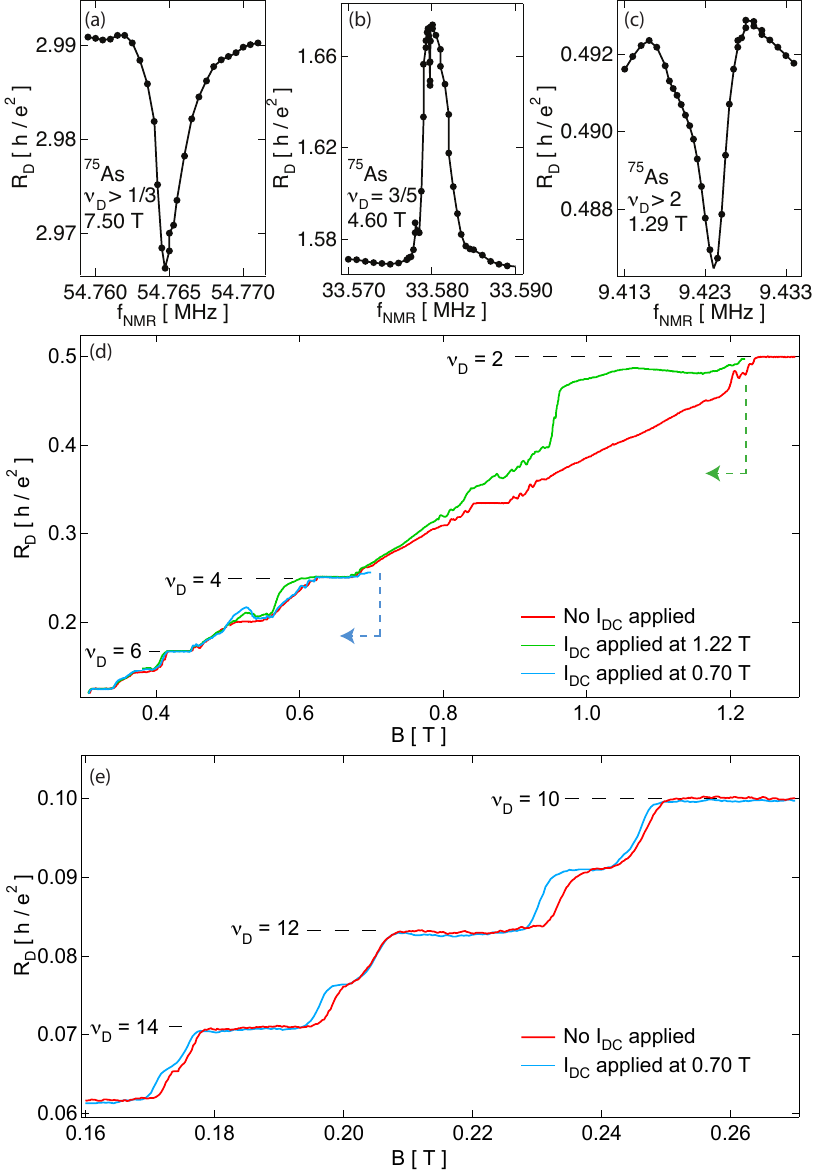}
\caption{\footnotesize{(a-c) $\rd$ as a function of $\fnmr$ in the $1~\um$ constriction at $B=7.50$~T,~4.60~T,~and~1.29~T. \cite{Detail1} (d,e) $\rd$ as a function of $B$ in the $1~\um$ constriction. Red trace indicates $\rd$ before $\Idc$ has been applied. Green trace indicates $\rd$ after $\Idc=54$~nA has been applied at $B=1.22$~T. Blue trace indicates $\rd$ after $\Idc=95$~nA has been applied at $B$~=~0.70~T.}}
\end{figure}

Following DNP at $\vqpc=3/5$, ramping the field to the edge of the $\vqpc=1$ plateau causes all plateaus to return to their unpolarized positions.  This rapid depolarization can be understood by the presence of skyrmions near $\nu=1$, which are known to cause relaxation of nuclear polarization \cite{Kuzma:2001p5872,Cata:1997p5729,Hashimoto:2002p4487}.  

At lower fields, plateau shifts and extensions in the IQH regime are seen following DNP from $\Idc$ applied between (not directly on) IQH plateaus. Following DNP, spin-split plateaus at $\vqpc=3$ and $\vqpc=5$ disappear for several minutes [Fig.~4(d)]. NMR confirms the DNP interpretation [Fig.~4(c)]. The sign and magnitude of the Overhauser field can also be deduced in the IQH regime. For $\nfield$ along $B$, ($\nfield>0$), spin splitting will always increase with DNP; for $\nfield$ opposing $B$, spin splitting will decrease for mild DNP, reach zero for $|\nfield|= B$ and again increase when $|\nfield|> B$ (with reversed spin splitting). Comparing Figs.~4(d,e), we see that for $B\sim 0.5-1$ T,  the odd (spin-split) plateaus are weakened by DNP, whereas for $B\sim 0.2$ T, odd plateaus are enhanced by DNP.    We conclude that $\nfield$ induced by DNP is directed opposite to $B$ and is between 0.2 and 0.5 T in magnitude. 

To connect DNP effects between IQH and FQH regimes, we polarize in the one regime and read out in the other. For instance, we apply $\Idc$ using the sweeping procedure at $\vqpc$~=~2, followed by ramping to a value of field where $\vqpc=3/5$ before polarization. Depolarization from skyrmions upon passing through $\vqpc = 1$ are avoided by fully depleting the QPC during the field ramp. We find that $\rd$ initially indicates $\vqpc=2/3$ value before sharply returning to the $\vqpc=3/5$ value after several minutes. Reversing the order---polarizing at $\vqpc=3/5$ and reading out at $2<\vqpc<3$---yields analogous results. We conclude from both procedures that the direction of induced Overhauser field opposes the applied field in both IQH and FQH regimes.  We also conclude that the relevant DNP occurs in the QPC (not downstream) since depolarization by skyrmions was eliminated by depleting only the electrons in the QPC constriction.

In the IQH regime, DNP presumably occurs by spin-up electrons at high-bias entering the QPC flipping into empty spin-down states, accompanied by a nuclear flop from spin down to spin up. Because dc bias exceeds Zeeman splitting but not cyclotron energy, the opposite mechanism, involving flip-flop spin relaxation between different Landau levels, which would would tend to align $\nfield$ and $B$, does not occur. 

Evidently, a similar mechanism appears to occur in the FQH regime. Within a CF picture, even filling factors can have spin-unpolarized ground states while odd filling factors are always at least partially polarized \cite{Jain07}. Hence, similar to electrons, CFs can be excited from a spin-up to a spin-down state.  Within this model, for example, exciting CFs from a spin-up subband of $\nu=3/5$ ($\vcf=-3$) to a spin-down subband of $\nu=4/7$ ($\vcf=-4$) will result in $\nfield<0$.  Excitations from spin-down to spin-up states may also be possible, however, since the CF Zeeman energy is comparable to the CF cyclotron energy \cite{Jain07}.

We interpret the effect of $\nfield$ on plateau structure as depending on ground-state spin configurations at successive filling factors. If successive states have different degrees of spin-polarization, $\nfield$ will change the length of the associated plateaus; if successive states are both spin-polarized, then $\nfield$ will shift plateau positions. In the IQH regime, odd filling factors are spin-polarized while even filling factors are spin-unpolarized, hence $\nfield$ causes plateaus at even filling factors to lengthen at the expense of plateaus at odd filling factors. In the FQH regime, the more spin-polarized state will also be destabilized by $\nfield$, leading to a shorter plateau.  We observe the destabilization of more spin-polarized plateaus in favor of less spin-polarized plateaus in both regimes: at $3/5\leq\vqpc\leq2/3~(-3\leq\vcf\leq-2)$ in Fig.~2(b) in the FQH regime and at $2\leq\vqpc\leq3$ in Fig.~4(d) in the IQH regime.  The changes in the lengths of the plateaus observed in Fig.~2(b) are analogous to those observed in Fig.~4(d), suggesting that CF's in the FQH regime are exhibiting the same behavior as electrons in the IQH regime.  Within this picture, shifts in plateau position can only occur in the FQH regime, where successive states can both be spin-polarized.  A change in Zeeman energy will not affect the size of the gap between spin-polarized CF Landau levels but will shift the energies of the levels equally.  When the Zeeman energy is decreased, the energy of each spin-polarized level increases, causing a local depopulation of electrons in the QPC.  Each energy level will then be filled at a lower magnetic field; the start of each plateau will then appear at a lower field than before DNP. We observe this shifting of the plateaus at $1/3<\nu<2/5$ [Fig.~2(a)].  Finally, while DNP is found to readily occur at $\vqpc=1/2$, it leaves little or no signature in $\rd$ at $\vqpc=1/2$, by symmetry, but can be observed by moving to another filling factor after DNP, as seen in Fig.~3(b).     

We thank J.~P.~Eisenstein, B.~I.~Halperin, I.~Neder, M.~S.~Rudner, K.~von Klitzing  for enlightening discussions.  Research funded by Microsoft Corporation Project Q, IBM, NSF (DMR-0501796) and Harvard University.  Device fabrication at Harvard's Center for Nanoscale Systems.

\small
\bibliography{DNP}
\end{document}